\documentclass[intlimits,twoside,a4paper]{article}

\usepackage{amsmath,amssymb}
\usepackage{graphicx}
\usepackage{wrapfig}
\usepackage{amssymb}
\usepackage{amsmath}

\usepackage[T2A]{fontenc}
\usepackage[cp1251]{inputenc}
\usepackage[eqsecnum]{cmpj2}

\issue{2012}{15}{1}{13704}

\doinumber{10.5488/CMP.15.13704}

\newcommand{\eps}{\varepsilon}

\title[High-frequency longitudinal oscillations]%
{High-frequency longitudinal oscillations\\ of quasi-two-dimensional electron liquid%
}
\author[V.M.~Gokhfeld, O.V.~Kirichenko, V.G.~Peschansky]{V.M.~Gokhfeld\refaddr{label1},
        O.V.~Kirichenko\refaddr{label2}, V.G.~Peschansky\refaddr{label2} }

\authorcopyright{V.M.~Gokhfeld, O.V.~Kirichenko, V.G.~Peschansky, 2012}
\addresses{
\addr{label1} Donetsk Institute for Physics and Engineering named
after O.O.~Galkin of the National Academy \\of Sciences of Ukraine, Donetsk,
Ukraine
 \addr{label2} B.I.~Verkin Institute for Low
Temperature Physics and Engineering of the National Academy \\of
Sciences of Ukraine, Kharkiv, Ukraine  }
\date{Received August 29, 2011, in final form  January 30, 2012}

\begin{document}

\maketitle

\begin{abstract}
The specific character of longitudinal collective electromagnetic
oscillations in a layered conductor with the quasi-two-dimensional
electron energy spectrum has been analyzed.
\pacs 71.20.-r
\end{abstract}

\section{Introduction}

Numerous low-dimensional conductors differ by their layered structure
and highly anisotropic electrical conductivity. Their in-plane
conductivity is much higher than that in perpendicular direction,
i.e. along the $OZ$-axis. Such an anisotropy is  typical of
cuprates, e.g. YBaCuO in a nonsuperconducting phase,
transition metal dichalcogenides (NbSe$_{2}$, TaS$_{2}$),
graphite and its intercalates in particular, as well as a broad
family of tetrathiafulvalene salts of (BEDT-TTF)$_{2}$I$_{3}$ type
\cite{1,2,3,4,5}. A common feature of many layered conductors is
their weak electron-energy dependence on the momentum projection
$p_{z}$. This is due to the fact that their interlayer distance
$a$ is much higher than the in-plane crystal lattice parameter.
Then, the wave functions of appropriate electrons  weakly overlap, and
the energy of single-particle charged excitations may be
represented by
rapidly converging series
\begin{equation}\label{1.1}
\varepsilon(\mathbf{ p})=\sum_{n=0}^{\infty}\varepsilon_{n}(\mathbf{
p}_{\bot})\cos\frac{anp_{z}}{\hbar}\, ,
\end{equation}
where the functions $\eps_{n}$ decrease with  increasing $n$
and $\varepsilon_{1}(\mathbf{ p}_{\bot})$ is much less than
$\varepsilon_{0}(\mathbf{ p}_{\bot})$. Here $\hbar$ is the Planck
constant, $\mathbf{ p}_{\bot}=[\mathbf{ pn}]$ is the momentum projection
onto the layer plane.

In conductors with  high charge carriers density allowance for the
electron-electron interaction is of great importance. The system
of quasiparticles carrying a charge should be regarded as the
Fermi liquid. According to the Landau-Silin theory~\cite{6,7}, the
interaction between quasiparticles may be taken into account in the
form of self-consistent field. In this case, the quasiparticle
energy depends on the distribution function $n(\mathbf{ p},\mathbf{ r},t)$
for the other quasiparticles. As a result, the energy of an
electron acquires the correction $\delta\eps$ to the
dispersion law~\eqref{1.1}
\begin{equation}\label{1.2}
\delta\varepsilon=\int\frac{2\rd^{3}p'}{(2\pi\hbar)^{3}}\ L(\mathbf{
p},\mathbf{ p}')n_{1}(\mathbf{ p}',\mathbf{ r},t),
\end{equation}
where $n_{1}$ is the correction to the equilibrium Fermi
distribution function $n_{0}(\varepsilon)$. In
quasi-two-dimen\-sio\-nal conductors, the Landau correlation function
$L(\mathbf{ p},\mathbf{ p}')$, as well as the quasiparticle energy in
the ``gas'' approximation~\eqref{1.1}, depends weakly on $p_{z}$.

The specificity of quasi-two-dimensional electron energy spectrum
provides the most favorable conditions for a detailed study of the
electron-electron Fermi liquid correlations. If the initial
dispersion law~\eqref{1.1} is unknown, investigation of nonequilibrium
processes in the electron system in a stationary  and uniform
external field does not provide novel information about
electron-electron Fermi liquid correlations. The solution of the
inverse problem of  restoring  the electron energy spectrum from
experimental data makes it possible to determine the dependence of the
electron energy $\widetilde{\varepsilon}(\mathbf{ p})=\varepsilon(\mathbf{
p})+\delta\varepsilon(\mathbf{ p})$ renormalized by the Fermi liquid
interaction, at the Fermi surface only.

However, wave processes are very sensitive to the form and
magnitude of the Fermi liquid correlations.  The gas approximation does not provide for propagation of the spin waves in nonmagnetic metals, predicted
by Silin~\cite{8}, as well as for a series of Fermi liquid effects. Spin waves and transverse
electromagnetic waves in the Fermi liquid have been studied in
detail not only in metals (see, for example, the
monograph~\cite{9,10}) but also in low-dimensional conductors
\cite{11,12,13,14,15,16,17}.

Here we consider longitudinal oscillations in the
quasi-two-dimensional Fermi liquid. Following Vla\-sov~\cite{18} and
Landau~\cite{19} we should solve the set of equations including the
Poisson equation
\begin{equation}\label{1.3}
\mathrm{div}\,\mathbf{ E}=4\pi\rho',
\end{equation}
the continuity equation
\begin{equation}\label{1.4}
\frac{\partial\rho'}{\partial t}+ \mathrm{div}\,\mathbf{ j}=0
\end{equation}
and the constitutive equation relating the electric field density
$\bf J$ to the electric field  $E\sim\exp(-\ri\omega t)$ with
allowance for the Fermi liquid correlations.

For the sake of brevity in calculations, we shall confine ourselves
to the simplest case where the electron dispersion law contains
only zero and first harmonics and the expression~\eqref{1.1} takes the
following form:
\begin{equation}\label{1.5}
\varepsilon(\mathbf{ p})=\varepsilon_{0}(\mathbf{
p}_{\bot})+t_{\bot}\cos\theta.
\end{equation}
Here $\varepsilon_{0}(\mathbf{ p}_{\bot})$ is an arbitrary function,
$t_{\bot}$ is the constant value, $\theta = {ap_{z}}{\hbar}^{-1}$.

We consider the case where the electric field and the wave vector
$\bf k$ are parallel to the normal to the layers, i.e. to the hard
direction for the electric current. We shall derive the dielectric
function with allowance for the Fermi liquid interaction (FLI),
ascertain the dispersion law for longitudinal plasma oscillations
and find out the distribution of the high-frequency electric field
in a half-infinite sample. We shall also consider low-frequency
collective excitations in a layered conductor which are possible
in the presence of two bands of the type~\eqref{1.5} in its electron
spectrum.

\section{ Dielectric function}

In order to determine the relation between the current density
\begin{equation}\label{2.1}
\mathbf{ j}=\int\ \frac{2\rd^{3}p}{(2\pi\hbar)^{3}}\ e\
\frac{\partial\tilde{\varepsilon}}{\partial\mathbf{ p}}\ n(\mathbf{
p},\mathbf{ r},t)
\end{equation}
and the electric field of the wave, it is necessary to solve the
kinetic equation
\begin{equation}\label{2.2}
\frac{\partial n }{\partial t }+\frac{\partial n }{\partial\mathbf{ r
}}\ \frac{\rd\mathbf{ r}}{\rd t}+\frac{\partial n}{\partial \mathbf{ p}}\
\frac{\rd \mathbf{ p}}{\rd t}=\hat{W}_{\mathrm{coll}}\{n\}\,.
\end{equation}
Naturally, the collision operator $\hat{W}_{\mathrm{coll}}\{n\}$ vanishes
working on  the equilibrium Fermi function
\begin{equation}\label{2.3}
n_{0}(\tilde{\varepsilon})=\left[1+\exp\left(\frac{\tilde{\varepsilon}-\mu}{T}\right)\right]^{-1}
\end{equation}
which depends on the energy $\tilde{\varepsilon}=\varepsilon(\mathbf{
p})+\delta\varepsilon(\mathbf{ p},\mathbf{ r},t)$ renormalized by the
Fermi liquid correlations.
 In the case of a weak perturbation of
the system of charge carriers, i.e. when the value of the electric
field of the wave is small, the collision integral is the linear
integral operator working on the function
\begin{equation}\label{2.4}
n_{1}(\mathbf{ p},\mathbf{ r},t)=n(\mathbf{ p},\mathbf{
r},t)-n_{0}(\tilde{\varepsilon})=-\Phi(\mathbf{ p},\mathbf{ r},t)
\frac{\partial
n_{0}(\tilde{\varepsilon})}{\partial\tilde{\varepsilon}}\,.
\end{equation}

We shall confine ourselves to the $\tau$-approximation for the
collision integral, where $\hat{W}_{\mathrm{coll}}$ corresponds to the
operator of multiplication nonequilibrium part of the distribution
function for conduction electrons by the the frequency of their
collisions $1/\tau$:
\begin{equation}\label{2.5}
\hat{W}_{\mathrm{coll}}\{n\}=\frac{1}{\tau}\ \Phi(\mathbf{ p},\mathbf{ r},t)\
\frac{\partial n
_{0}(\tilde{\varepsilon})}{\partial\tilde{\varepsilon}}\,.
\end{equation}

In the kinetic equation, it should be taken into account that
besides the electric field, conduction electrons experience the
force of the self-consistent field of interacting quasiparticles
\begin{equation}\label{2.6}
\frac{\rd\mathbf{ p}}{\rd t}=\mathbf{ F}=e\mathbf{ E}-\frac{\partial }{\partial\mathbf{
r}}\ \delta\varepsilon(\mathbf{ p},\mathbf{ r},t).
\end{equation}

 The current density depends only on the nonequilibrium part of
the distribution function of charge carriers
\begin{equation}\label{2.7}
\mathbf{ j}(\mathbf{ r},t)=-e\int\ m^{*}\rd\varepsilon\int \rd\theta\int
\rd\varphi  \frac{\partial n _{0}(\varepsilon)}{\partial\varepsilon
}\mathbf{ v}\Phi(\varepsilon,\theta,\varphi,\mathbf{
r},t)\frac{1}{2\pi^{3}\hbar^{2}a}\,.
\end{equation}
Here $m^{*}$ is the cyclotron effective mass, $\varphi$ is the
variable of integration over charge carriers states with constant
values of energy and momentum projection on the normal to the
layers, $\mathbf{ v} = {\partial\varepsilon(\mathbf{ p}) }/{\partial
\mathbf{ p}}$.

 The kinetic equation for the case
under consideration takes the form
\begin{equation}\label{2.8}
  kv_{z}\Phi-\omega\Psi+\ri e Ev_{z}=\ri\frac{\Phi}{\tau}\, ,
\end{equation}
where $\Psi = \Phi - \delta\varepsilon$, the relaxation time $\tau$ is supposed to be sufficiently
large.

Within the chosen model for energy spectrum~\eqref{1.5}, electron
velocity across the layers $v_{z} = v_{1} \sin ( \theta )$, with
$v_{1} = - a {t_{\bot}}{\hbar}^{-1}$, is dependent on $\theta$ only
and much less than the characteristic Fermi velocity $v_{\mathrm{F}}$ for
the motion along the layers. Then, in order to find out the
dielectric permeability and finally the spectrum of longitudinal
vibrations, the averaging over $\phi$ of the nonequilibrium
distribution function of the charge carriers along with the
Lanadau-Silin correlation function is sufficient.

In accordance with  the symmetry of the problem, the Landau
correlation function $L(\mathbf{ p},\mathbf{ p}')$, which connects the
effective and the real distribution of charge carriers, may be
represented in the form
\begin{equation}\label{2.9}
L(\theta,\theta')=L_{0}+2L_{1}\cos(\theta-\theta').
\end{equation}
In this case
\begin{equation}\label{2.10}
\Phi(\theta)=\Psi(\theta)+\int_{-\pi}^{\pi}\
\frac{\rd\theta'}{2\pi}\ L(\theta,\theta')\Psi(\theta')
\end{equation}
and one can easily solve the equations~\eqref{2.8}--\eqref{2.10}. In particular,
for $\Psi_{0}=\langle\Psi\rangle/\langle1\rangle$, averaging over
$\theta$ we have
\begin{equation}\label{2.11}
\Psi_{0}=\frac{\ri e E}{k}\ \frac{\tilde{\omega}
W}{\omega+W\left[\ri\tau^{-1}+\tilde{\omega}\left(L_{0}+\lambda\omega^{2}/k^{2}v_{0}^{2}\right)\right]}\, ,
\end{equation}
where
\begin{equation}\label{2.12}
W(k,\omega)=\left\langle\frac{kv_{z}}{kv_{z}+\tilde{\omega}}\right\rangle\langle
1\rangle^{-1}=1-\frac{\tilde{\omega}}{\sqrt{\widetilde{\omega}^{2}-k^{2}v_{1}^{2}}}\,,
\end{equation}
\[
\lambda=\frac{2L_{1}}{(1+2L_{1})}\,,\qquad  \tilde{\omega}=\omega+\frac{\ri}{\tau}\,,\qquad   \langle 1\rangle=\frac{2}{(2\pi\hbar)^{3}}
\ \int\frac{\partial n_{0}}{\partial\varepsilon}\rd^{3}p \, .
\]

With regard to~\eqref{2.7} and~\eqref{2.8}, the dielectric function
$\epsilon=1+4\pi\langle\Psi\rangle/\ri kE$ takes the form
\begin{equation}\label{2.13}
\epsilon(k,\omega)=1+\frac{\kappa^{2}}{k^{2}}\
\frac{\tilde{\omega}
W}{\omega+W\left[\ri\tau^{-1}+\tilde{\omega}\left(\frac{L_{0}+\lambda\omega^{2}}{k^{2}v_{0}^{2}}\right)\right]}\, ,
\end{equation}
where
\[ \kappa^{2}=4\pi e^{2}\langle
1\rangle=\frac{4m^{2}e^{2}}{a\hbar^{2}}\]
is the square of the decrement
of the static screening in the gas approximation. As is easily
seen from \eqref{2.13} and the dispersion equation
$\epsilon(k,\omega)=0$, the decrement with allowance for FLI
equals $\kappa\left(\sqrt{1+L_{0}}\right)^{-1}$. For the activation frequency of
the plasma oscillations we have
\begin{equation}\label{2.14}
\omega_{\mathrm{p}}^{2}=\frac{2m e^{2}v_{1}^{2}}{a\hbar^{2}}\
(1+L_{1})=\eta\ \frac{4\pi N e^{2}}{m}(1+L_{1})\, ,
\end{equation}
where $\eta=(v_{1}/v_{\mathrm{F}})^{2}$. Due to the smallness of $\eta$, the
plasma frequency in the quasi-two-dimensional conductor essentially reduces
  in comparison with the case of a conventional metal.
 Making use of the formula~\eqref{2.14} we obtain
the dispersion law for longitudinal plasmons:
\begin{eqnarray}\label{2.15}
\omega^{2}(k)&=&\frac{\omega_{\mathrm{p}}^{2}}{\lambda}\ \left[\
\frac{k^{2}}{2\kappa^{2}}-K+\lambda
K+\sqrt{\left(K-\frac{k^{2}}{2\kappa^{2}}\right)^{2}+\lambda
K\frac{k^{2}}{\kappa^{2}}}\ \right],
\\\nonumber
K(k)&=&1+\frac{(1+L_{0})k^{2}}{\kappa^{2}}\,.
\end{eqnarray}
When the Landau correlation function has only a zero harmonic, the
dispersion relation becomes noticeably simpler:
\begin{equation}\label{2.16}
\omega(k)=\frac{e v_{1}}{\hbar}\sqrt{\frac{2m}{a}}\ \
\frac{\kappa^{2}+(1+L_{0})k^{2}}{\kappa\sqrt{\kappa^{2}+(1/2+L_{0})k^{2}}}\,.
\end{equation}

\section{Distribution of the electric field in a specimen}

Having calculated the dielectric function for the infinite
specimen, we can solve the boundary problem for the half-space
$z\geqslant  0$. Landau~\cite{19} was the first to consider the penetration of the high-frequency
longitudinal electric field into the Maxwell plasma. Propagation of the longitudinal
waves in a degenerated isotropic conductor was studied in
\cite{20} where a detailed analysis of the boundary conditions
for an electron distribution function was  made. In the
layered conductor, the angles of incidence of electrons on the
boundary do not exceed $v_{1}/v_{\mathrm{F}}\ll1$, which allows us to suppose
that charge carriers are reflected specularly. The electric field
 outside the conductor (i.e. between the capacitor plates, one of
which is the specimen) is supposed to be given by $E(z\leqslant
0)=(0,0,E_{0}\exp(-\ri\omega t))$. Then, the field distribution in
the specimen is described by the following expression
\begin{equation}\label{3.1}
E(z)=\frac{E_{0}}{\ri\pi}\ \int_{-\infty}^{\infty}\
\frac{\rd k}{k\epsilon(k,\omega)}\exp(\ri kz).
\end{equation}
The $k=0$ pole residue gives the field value in the bulk of the
conductor $E(\infty)$ which is much less than $E_{0}$, if
$\omega\ll\omega_{\mathrm{p}}\,$:
\begin{equation}\label{3.2}
E(\infty)=\frac{E_0}{\epsilon(k,\omega)}\,,\qquad
\epsilon(0,\omega)=1-\frac{\omega_{\mathrm{p}}^{2}}{\omega(\widetilde{\omega}+\ri L_{1}/\tau)}\,.
\end{equation}
However, at $\omega\tau\gg 1$,  the field attenuates in a
nonmonotonous manner because the integral~\eqref{3.1} contains an
oscillating component
\begin{equation}\label{3.3}
E_{1}(z)= -\frac{4E_{0}F^{2}}{\pi} \int_{1}^{\infty}
\frac{\rd xx\sqrt{x^{2}-1}\exp(\ri k_{1}xz)}{\left(x^{2}-1\right)\left[2+(1+L_{0})F^{2}x^{2}\right]^{2}+\left(2+L_{0}F^{2}x^{2}\right)^{2}}
\end{equation}
originating from the branching point $k_{1}=\tilde{\omega}/v_{1}$.
At $z\gg v_{1}/|\tilde\omega|$, the ratio of the oscillating
component to $E(\infty)$ takes the form
\begin{equation}\label{3.4}
\frac{E_{1}(z)}{E_{\infty}}=\frac{1-F^{2}}{\sqrt{2\pi}(1+L_{0}F^{2}/2)^{2}}\
\left(\frac{v_{1}}{\tilde\omega z}\right)^{3/2}\ \exp(\ri z\tilde\omega/v_{1}).
\end{equation}
Here $F=\tilde\omega/\omega_{\mathrm{p}}$. In the layer of the order of the
electron free path $l_{z}=v_{1}\tau$, the field oscillates with
the period $2\pi v_{1}/\omega$. In the layered conductor, both
scale lengths reduce in comparison with the isotropic metal,
remaining, however, microscopic for non-high frequencies $\omega$
and $\tau^{-1}$.

 For the sake of completeness, the resonance case $\omega=\omega_{\mathrm{p}}$
 should be considered. At  such high frequencies, the direct
stimulation of the monochromatic field is hard to achieve, but the
character of the spatial distribution of the resonance harmonics
may become apparent in the pulsed mode, as well as in the
experiments with electron beams.

Near the plasma frequency $\omega_{\mathrm{p}}$, the radicals in the
expression for the dielectric function may be expanded into a
series about small $k^{2}$, and  in the main approximation we have
\begin{equation}\label{3.5}
\epsilon(0,\omega)=1-\frac{\omega_{\mathrm{p}}^{2}}{\omega(\tilde\omega+\ri L_{1}/\tau)}\,,\qquad  |\epsilon(0,\omega)|\ll 1.
\end{equation}

\section{ The spectrum of the short-wave plasmons}

Consider the case of large values of $k$ in collisionless limit.
Strictly speaking, this is the case where the quasiclassical
formula for the conductivity
\begin{equation}\label{4.1}
\sigma(\mathbf{ k},\omega)=-\ri e^{2}\left\langle\frac{v_{z}^{2}}{\mathbf{ kv
}-\tilde\omega}\right\rangle
\end{equation}
is inappropriate because it is not invariant with respect to
translation  $k\rightarrow k + 2\pi /a$.  The
translation-invariant dielectric function $\epsilon(k)=1+4\pi
\ri\sigma(k)/\omega$ can be easily constructed if we note that the
magnitude $\hbar\mathbf{ kv}$ is the result  of the expansion in small
$\bf k$ of the energy difference in the quantum perturbation
theory formulas. In the case of the spectrum~\eqref{1.5}, this difference
is
\[
\varepsilon\left(\mathbf{ p}+\hbar\frac{\mathbf{ k}}{2}\right)-\varepsilon\left(\mathbf{ p}-\hbar\frac{\mathbf{ k}}{2}\right)=\hbar v_{z}\left(\frac{2}{a}\right)\sin\left(k\frac{a}{2}\right)\, ,
\]
i.e., in~\eqref{4.1} $k$ should be replaced by $Q=(2/a)\sin(ka/2)$. As a
result, we have
\begin{equation}\label{4.2}
\epsilon(k,\omega)=1+\frac{\kappa^{2}}{Q^{2}}\left(1-\frac{\omega}{\sqrt{\omega^{2}-Q^{2}v_{1}^{2}}}\right).
\end{equation}
Usually, the application of the quasiclassical approximation for the
calculation of the plasmon spectrum is limited by the condition
$\hbar k\ll p_{\mathrm{F}},\hbar/a$ (see, for example,~\cite{21}). In the
case under consideration, this restriction is of no necessity
because the dispersion law~\eqref{1.5} is determined everywhere in the
Brillouin zone. From~\eqref{4.2} we obtain:
\begin{equation}\label{4.3}
\omega(k)=\omega_{p0}\ \sqrt{\frac{2}{b}}\
\frac{b+\sin^{2}(ka/2)}{\sqrt{2b+\sin^{2}(ka/2)}}\, ,
\end{equation}
where $\omega_{p0}=(ev_{1}/\hbar)\sqrt{2m/a},\ b=(\kappa
a)^{2}/4=ane^{2}/\hbar^{2}$ i.e., the ratio of the lattice period
to the Bohr radius. As far as Bohr radius (at $m\simeq m_{0}$) is
of the order of $10^{-8}$ cm, the parameter $b$ may be very
large, especially in artificial superlattices with great
separation between conducting layers. In this case, the spectrum
\eqref{4.3} corresponds to a narrow band
\begin{equation}\label{4.4}
\frac{\omega_{\mathrm{p}}^{\mathrm{max}}}{\omega_{0}}-1=\frac{b+1}{\sqrt{b^{2}+b/2}}-1\simeq
\left\{\begin{array}{c}
  3/4b\ \ (b\gg 1) \\
  \sqrt{2/b}\ \ (b\ll 1)
\end{array}
\right.
\end{equation}
and the static screening decrement $\kappa$ is replaced by
\begin{equation}\label{4.5}
\overline{\kappa}=\frac{2}{a}\mathrm{arsinh} \left(\sqrt{b}\right).
\end{equation}

It is easily seen that the results obtained above are valid in an
external magnetic field  applied across the layers, because it has
no effect on the electron velocity along the normal to the layers
for the energy spectrum under discussion.

\section{Two-band model}

The Fermi surface is supposed to be singly connected.
However, the electron structure of ``synthetic metals'' is quite
complicated. It is probable that there exist two charge carrier
groups (for example, electrons and holes) of the type~\eqref{1.5} with
the different values $V_{z\mathrm{max}} :v_{1}^{(2)}=Bv_{1}^{(1)}$ and
$B\geqslant  1$.
  In the case of
two charge carrier groups  in an isotropic metal FLI (even in its
simplest form $L(\mathbf{ p}, \mathbf{ p}')=L_{\alpha,\beta},\
\alpha,\beta, = 1, 2 )$, there appears a longitudinal
collective mode of the type of zero sound at sufficiently low
frequencies, $\tau^{-1}\ll\omega\ll\omega_{\mathrm{p}}$. Its phase velocity
$V =uv_{1}^{(2)}$ satisfies the characteristic equation
\begin{equation}\label{5.1}
D(k)=\sum\zeta_{\alpha}W_{\alpha}=LW_{1}W_{2}=\zeta_{1}W(Bu)+\zeta_{2}W(u)+LW(Bu)W(u)=0,
\end{equation}
where
\[
\zeta_{\alpha}=\frac{m_{\alpha}}{m_{1}+m_{2}}\,,\qquad  L=\zeta_{1}\zeta_{2}\left(L_{11}+L_{22}-L_{12}-L_{21}\right)
\]
and $W(u)=W(\tilde\omega/kv_{1}^{(2)})$  is given by the formula
\eqref{2.12}. The equation~\eqref{5.1} has a real root $(u>0)$ only at sufficiently
intensive FLI, namely at \[L\geqslant
L_{\mathrm{min}}(B)=-\zeta_{2}/W(B).\]  In the quasi-two-dimensional metal,
the threshold value $L_{\mathrm{min}}$ is much less than that in the
isotropic metal for the same $B$, especially near $B=1$. In this
case  $L_{\mathrm{min}}=0$, and the solution of the equation~\eqref{5.1} is
\begin{equation}\label{5.2}
u(L)=\frac{1+L}{\sqrt{1+2L}}\,.
\end{equation}

The wave under consideration corresponds to antiphase partial
oscillations of the electron densities for both charge carrier
groups. The elastic force generated by non-equilibrium charge
carrier distribution allows one to detect the electron zero sound
by means  of the concomitant elastic wave. This effect was
observed in a series of ordinary metals (W, Al, Ga)~\cite{21}. It
may be described by jointly solving the electron kinetic equation
and the elastic theory equation for the half-space with the given
oscillation of the boundary $u_{0}\exp(-\ri\omega t)$. As a result,
for the displacement of  ions in the $k$-representation, we obtain
\begin{eqnarray}
U(k)&=&\frac{2U_{0}}{\ri k}\left[1+\frac{\omega^{2}}{s^{2}k^{2}(1+R)-\omega^{2}}\right],
\\
\nonumber
\label{5.3}
R(k)&=&\frac{\omega\Lambda^{2}\langle 1 \rangle}{\tilde\omega\rho
s^{2}}\ \zeta_{1}\zeta_{2}(1+L) \left[\ 1-W_{1}W_{2}\
\frac{1+L}{D(k)}\ \right],
\end{eqnarray}
where  $\rho$ is the mass density, $s$ is the velocity of the
longitudinal sound. We assume that $s\ll v_{1}$. In this case,
$U(k)$ has two different poles: the pole $k_{s}\simeq\omega/s$ is
connected to the ordinary sound; the other one,  $k_{1s}$, is
close to the root of the equation~\eqref{5.1} and describes an extra
elastic wave. It is easy to see that in the degenerated case
$(v_{1}^{(1)}= v_{1}^{(2)}=v)$ we have:
\begin{equation}\label{5.4}
U_{0s}\simeq U_{0}\ \frac{\omega\Lambda^{2}\langle 1
\rangle}{\tilde\omega\rho v_{1}^{2}}\ \zeta_{1}\zeta_{2}\
\frac{2L}{1+L}\ \exp\left[\frac{\ri\tilde\omega
z\sqrt{1+2L}}{v_{1}(1+L)}\right].
\end{equation}
This expression is proportional to $v_{1}^{-2}$  instead of
$v_{\mathrm{F}}^{-2}$ in an ordinary metal.

\section{Conclusions}

The analysis given above shows that the electrodynamic
characteristics of the layered conductor essentially differ from
those of the isotropic metal with the same charge carrier
density. In particular, the activation frequency and the velocity
of the longitudinal waves, propagating along the normal to the
layers, decrease essentially. Moreover, the case of
quasi-two-dimensional charge carrier spectrum is most favorable
in observing the electron zero sound and in studying the
electron phenomena in various modifications of carbon.

\ukrainianpart

\title{Високочастотні поздовжні осциляції квазі-двовимірної електронної рідини}
\author{В.М. Гохфельд\refaddr{label1,label2}, О.В. Кириченко\refaddr{label2}, В.Г. Піщанський\refaddr{label2}}

\addresses{
\addr{label1} Донецький фізико-технічний інститут ім.~А.А. Галкіна, Донецьк, Україна %
\addr{label2} Фізико-технічний інститут низьких температур ім.~Б.І. Вєркіна НАН України, Харків, Україна}

\makeukrtitle

\begin{abstract}
Проаналізовано специфіку поздовжніх колективних електромагнітних коливань у шаруватому провіднику з квазі-двовимірним електронним енергетичним спектром.%
\keywords шаруваті провідники,  фермі-рідина, поздовжні коливання

\end{abstract}
\end{document}